\documentclass{emulateapj}

\def\src{4U 1626$-$67}
\def\uhuru{{\sl Uhuru}}

\def\asca{{\sl ASCA}}
\def\bepposax{{\sl BeppoSAX}}

\def\einstein{{\sl Einstein}}

\def\chandra{{\sl Chandra}}
\def\xmm{{\sl XMM-Newton}}

\def\ergcms{\hbox{erg cm$^{-2}$ s$^{-1}$ }}
\def\ergcmkev{\hbox{erg cm$^{-2}$ s$^{-1}$ keV$^{-1}$}}

\def\it{\sl}

\begin{document}

\title{High-resolution X-ray spectroscopy of the ultracompact LMXB pulsar 4U 1626--67}

\author{Miriam I. Krauss\altaffilmark{1}, Norbert S. Schulz, and Deepto Chakrabarty\altaffilmark{1}}
\affil{Kavli Institute for Astrophysics and Space Research, Massachusetts Institute of Technology,
Cambridge, MA 02139 \\
miriam, nss, deepto@space.mit.edu}
\altaffiltext{1}{Also Department of Physics, Massachusetts Institute of Technology, Cambridge, MA 02139}

\author{Adrienne M. Juett}
\affil{Department of Astronomy, University of Virginia, Charlottesville, VA 22903 \\
ajuett@virginia.edu}

\author{and \\
Jean Cottam}
\affil{Laboratory for High Energy Astrophysics, NASA Goddard Space Flight Center, Greenbelt, MD, 20771 \\
jcottam@milkyway.gsfc.nasa.gov}


\begin{abstract}

We report results from four recent observations of the ultracompact low-mass X-ray binary pulsar 4U 1626$-$67.  All the observations obtained high-resolution X-ray spectra of the system, two from the {\it Chandra X-ray Observatory} using the High Energy Transmission Grating Spectrometer, and two from the {\it XMM-Newton Observatory} using the Reflection Grating Spectrometer as well as the EPIC PN and MOS.  These data allow us to study in detail the prominent Ne and O emission line complexes which make 4U 1626$-$67 unique among LMXBs.  The observations were spaced over a period of 3 years for a total observing time of 238 ks, allowing us to monitor the line regions as well as the overall source flux, continuum spectrum, and timing properties.  The structure of the emission lines and the ratios of the components of the helium-like \ion{Ne}{9} and \ion{O}{7} triplets support the hypothesis that they are formed in the high-density environment of the accretion disk.  We do not find any significant changes in the line widths or ratios over this time period, though we note that the line equivalent widths decrease.  Using the most recent calibration products, we are able to place constraints on the strengths of the Ne K, Fe L, and O K photoelectric absorption edges.  In contrast to our earlier analysis, the data do not require an overabundance of Ne or O in the system relative to the expected ISM values.  We find that the pulsar is still spinning down, though the rate of decrease is greater than what was predicted by the ephemeris derived from previous {\it Compton}/BATSE monitoring.  We also note that the pulse profile has changed significantly from what was found prior to the torque reversal in 1990, suggesting that this event may be linked to a change in the geometry of the accretion column.  The flux of 4U 1626$-$67 continues to decrease, in keeping with the trend of the last $\approx 30$ years over which it has been observed.  Taking into consideration current theory on disk stability, we expect that 4U 1626$-$67 will enter a period of quiescence in 2--15 years.

\end{abstract}

\section{Introduction}

The 7.7 s X-ray pulsar \src~was first discovered by \uhuru~\citep{giacconi72,rappaport77}, and remains the only known  high-field \citep[$B \approx 4 \times 10^{12}$~G,][]{pravdo79,coburn02} pulsar in an ultracompact low-mass X-ray binary (LMXB).  This unique pairing of an apparently young neutron star with a very low-mass companion could indicate that the neutron star was originally a white dwarf whose mass, as a result of accretion, exceeded the Chandrasekhar limit \citep[][though see Verbunt et al. 1990]{joss78}. Although orbital motion has never been detected in X-ray data, pulsed optical emission reprocessed on the surface of the secondary allowed \citet{middleditch81} to infer an orbital period of 42 min, which was later confirmed by \citet{chakrabarty98a}.  It is therefore a member of the class of objects known as ``ultracompact" binaries ($P_{\rm orb} < 80$ min) which must have hydrogen-depleted secondaries to reach such short periods \citep{paczynski81,nelson86}.  It has an extremely small mass function of $f \leq 1.3 \times 10^{-6}~M_{\sun}$, which corresponds to a secondary mass of $0.04~M_{\sun}$ for $i = 18\degr$ \citep{levine88}.  A very low-mass secondary would account for the faint ($V \approx 17.5$) optical counterpart and the high optical pulsed fraction \citep{mcclintock77,mcclintock80}.

Initially, \src~was observed to be spinning up with a characteristic timescale $P/\dot{P} \approx 5000~\rm yr$, but in 1990, this trend reversed and the neutron star began to spin down on approximately the same timescale \citep{wilson93,chakrabarty97}.  The torque reversal was abrupt, although the decrease in bolometric X-ray flux has been gradual and continuous over the past $\approx 30$ yr.  Assuming that the X-ray flux is a good proxy for the accretion rate onto the neutron star, these observational facts cannot be reconciled with current accretion disk-neutron star interaction theory.  However, the fact that the pulsar underwent torque reversal implies that the radius at which the accretion disk is truncated by the neutron star's magnetic field is close to the corotation radius.  With this assumption and a neutron star mass of $1.4~M_{\sun}$, the measured $\dot{P}$ during spin-up can constrain the accretion rate to be $\gtrsim 2 \times 10^{-10}~M_{\sun}~\rm yr^{-1}$, which implies a distance of $\gtrsim 3$ kpc \citep{chakrabarty97}.  Alternatively, measurements of the optical and X-ray fluxes (assuming the albedo of the disk is $\gtrsim 0.9$) give a distance range to \src~of $5 \lesssim D \lesssim 13$ kpc \citep{chakrabarty98a}.

Perhaps the most unique characteristic of \src~is its X-ray spectrum.  \citet{angelini95} first reported the presence of Ne and O emission lines in an \asca~spectrum, and they found similar features in an \einstein~observation performed prior to the torque reversal.  Using a subsequent observation by \chandra, we discovered double-peaked line structure indicative of formation in an accretion disk \citep[][hereafter S01]{schulz01}.  Emission line features such as these are not seen in any other LMXB systems, and suggest that the donor is particularly rich in elements resulting from later stages of nuclear burning---perhaps a C-O-Ne or O-Ne-Mg white dwarf.  A UV spectrum obtained with {\it HST}/STIS revealed both emission and absorption features from C, O and Si \citep{homer02}.  The C absorption lines are stronger than expected from a purely ISM contribution, suggesting some local contribution, and the \ion{O}{5} line has a pronounced double-peaked profile, indicating an accretion-disk origin similar to the X-ray lines.  The spectrum is missing common N and He lines typically seen in UV spectra from high-excitation systems, likely because \src~is lacking in these elements.  A high-S/N optical spectrum obtained by \citet{werner06} using the VLT confirms the lack of He in the system, and finds H lacking as well.  The optical spectrum is dominated by C and O emission lines, but does not appear to contain any Ne lines.  Given the current understanding of the model Ne atom, they find that Ne is present in \src~at $\lesssim 10\%$ by mass; or, equivalently, that Ne/O $\lesssim 0.2$.

We present a series of four high-resolution X-ray spectra spanning three years, two obtained with the High Energy Transmission Gratings (HETGS) aboard the {\it Chandra X-ray Observatory}, and two from the Reflection Grating Spectrometers (RGS) aboard {\it XMM-Newton}.  We previously presented analysis of the first \chandra~observation in S01, but this re-analysis uses improved calibration products and software.  We describe the observations and data reduction procedures in Section~\ref{obs_redux}, and our spectral analysis in Section~\ref{spec_anal}.  We present timing analysis is Section~\ref{timing_anal}.  Finally, we discuss the implications of our observations in Section~\ref{disc}.

\section{Observations and Data Reduction}\label{obs_redux}

\subsection{{\it Chandra}}

4U 1626--67 has been observed twice with the HETGS on board the {\it Chandra X-Ray Observatory} \citep{canizares05}, first on 2000 September 16 and again on 2003 June 3.  The HETGS comprises two sets of transmission gratings:  the Medium Energy Gratings (MEGs), with a spectral resolution of $\Delta\lambda=0.023$ \AA~FWHM and a range of 2.5--31 \AA~(0.4--5.0 keV), and the High Energy Gratings (HEGs), which have $\Delta\lambda=0.012$ \AA~FWHM and a range of 1.2--15 \AA~(0.8--10 keV).  See Table~\ref{observations} for a summary of the observations.

\begin{deluxetable*}{llll}
\tablewidth{0pt}
\tabletypesize{\small}
\tablecaption{X-Ray Observations}
\tablehead{
\colhead{Observatory} & \colhead{Observation start} & \colhead{Observation ID} & \colhead{Duration} 
}
\startdata
{\it Chandra} & 2000 September 16 14:57 UT & 104 & 40 ks \\
{\it XMM-Newton} & 2001 August 24 02:57 UT & 0111070201 & 17 ks \\
{\it Chandra} & 2003 June 3 02:30 UT & 3504 & 97 ks \\
{\it XMM-Newton} & 2003 August 20 05:55 UT & 0152620101 & 84 ks \\
\enddata
\label{observations}
\end{deluxetable*}

We obtained ``level 1'' event lists from the {\it Chandra} data archive\footnote{http://cda.harvard.edu/chaser/mainEntry.do}, and reprocessed the data using the latest available version of the CIAO software (v.3.3) and the calibration database (CALDB v.3.2.1).  This processing applied the gain, time-dependent gain, and charge transfer inefficiency (CTI) corrections.  The CTI corrections improve the computation of event grade (used to filter out likely non-source events), and the gain products improve the channel to energy mapping of events, which allows for more precise order-sorting of the dispersed spectra.  We created grating responses that include the time-dependent effects of a contaminant present on the ACIS optical blocking filter \citep{marshall04}.  Correcting for this contamination is necessary for precise spectral analysis, since it affects both spectral shape and normalization.

Neither {\it Chandra} observation contained any appreciable background flares.  After reprocessing and filtering out bad pixels, we applied the standard grade filtering (retaining grades 0, 2, 3, 4 and 6), and used the {\tt tgdetect} tool to determine the zeroth-order source position. The high source count rate caused the zeroth-order images to be quite piled up, but this did not hinder our ability to determine the source position.  This position was used to sort the dispersed spectra into the proper orders, resulting in 62555 (101799) background-subtracted events in the first-order MEG spectrum and 35343 (59706) in the first-order HEG spectrum for ObsID 104 (3504).  The dispersed spectra are not affected by pileup.

\subsection{{\it XMM-Newton} }

4U 1626--67 has been observed four times with {\it XMM-Newton}, but only two of these observations contain a significant amount of science data (ObsIDs 0111070201, performed 2001 August 24, and 0152620101, performed 2003 August 20; see Table~\ref{observations}).  Here, we analyze data from the three EPIC cameras (two MOS detectors and the PN), as well as the two reflection grating spectrometers (RGS1 and RGS2).  The MOS (PN) detectors have nominal bandpasses of 0.15--12 (0.15--15) keV, and spectral resolutions of 70 (80) eV at 1 keV.  The RGSs have bandpasses of 0.35--2.5 keV and first-order spectral resolutions of 0.04 \AA~FWHM.
Both observations were performed after the failure of RGS1's CCD7 and RGS2's CCD4, resulting in gaps in the dispersed spectra which were filtered out for analysis.

The {\it XMM-Newton} science products were obtained from the {\it HEASARC}\footnote{http://heasarc.gsfc.nasa.gov/db-perl/W3Browse/w3browse.pl} archive.   Both observations were affected by strong background flares in the MOS and PN detectors, which we filtered out prior to analysis, resulting in the loss of $\approx$ 5 ks of MOS data and $\approx$ 2 ks of PN data for ObsID 0111070201, and $\approx$ 25 ks of MOS and PN data and $\approx$ 12 ks of RGS data for ObsID 0152620101.

During ObsID 0111070201, the MOS1 detector was in timing mode, so was not easily susceptible to pileup.  However, the source core is somewhat piled in the MOS2 detector, and was excised prior to analysis.  During ObsID 0152620101, the MOS2 detector was in timing mode, and the MOS1 detector shows mild evidence of pileup, so we again removed the central portion of source data prior to analysis. The standard filters were applied, FLAG == 0 and PATTERN $<=$ 12 (4), to the MOS (PN) data.  The RGS data were reprocessed (using {\it XMMSAS} v.6.5.0) with refined source coordinates, resulting in updated source and background spectra as well as responses.  Background spectra were created for the EPIC data using off-source regions, and scaled appropriately to match the source spectra.  After processing, there were a total of $1.1 \times 10^{5}$ ($1.7 \times 10^{5}$) background-subtracted counts in the MOS1 data, $7.3 \times 10^{4}$ ($4.5 \times 10^{5}$) in the MOS2 data, $3.1 \times 10^{5}$ ($1.04 \times 10^{6}$) in the PN data, 8366 (17289) in the RGS1 spectrum, and 9951 (21076) in the RGS2 spectrum for ObsID 0111070201 (0152620101).

\section{Spectral Analysis}\label{spec_anal}

We used the ISIS software package\footnote{http://space.mit.edu/CXC/ISIS/} (v1.3.3) for spectral fitting.  All source and background data, as well as response files, were read into ISIS, and normalized background counts were subtracted prior to fitting.  We combined the $\pm$1 spectral orders of the HETGS data, and used the $-1$ order of the RGS data.  We employed Cash statistics throughout our analysis, and fit the data from different instruments simultaneously.  All the quoted errors are 90\% confidence limits unless otherwise noted.

\subsection{Continuum Fitting}

For continuum fitting, we excluded the known emission line regions, and used the wavelength/energy ranges 1.2--17.0 (1.8--26.0) \AA~for the HEG (MEG) grating spectra, 7.0--34.0 \AA~for the RGS grating spectra, and 1.1--9.9 (1.1--8.3) keV for the MOS (PN) spectra.  The lower energy ranges of the MOS and PN spectra were excluded because they contain unresolved Ne and O emission lines which artificially increase the overall continuum level.   We also excluded regions in the RGS data that contain bad columns or fall across non-functioning CCD chips.  During the second \xmm~observation (ObsID 0152620101), MOS2 was in timing mode, and its spectral calibration did not match that of the other instruments, so these data were omitted from the continuum fit.  

\begin{figure}
\resizebox{3.5in}{!}{\includegraphics{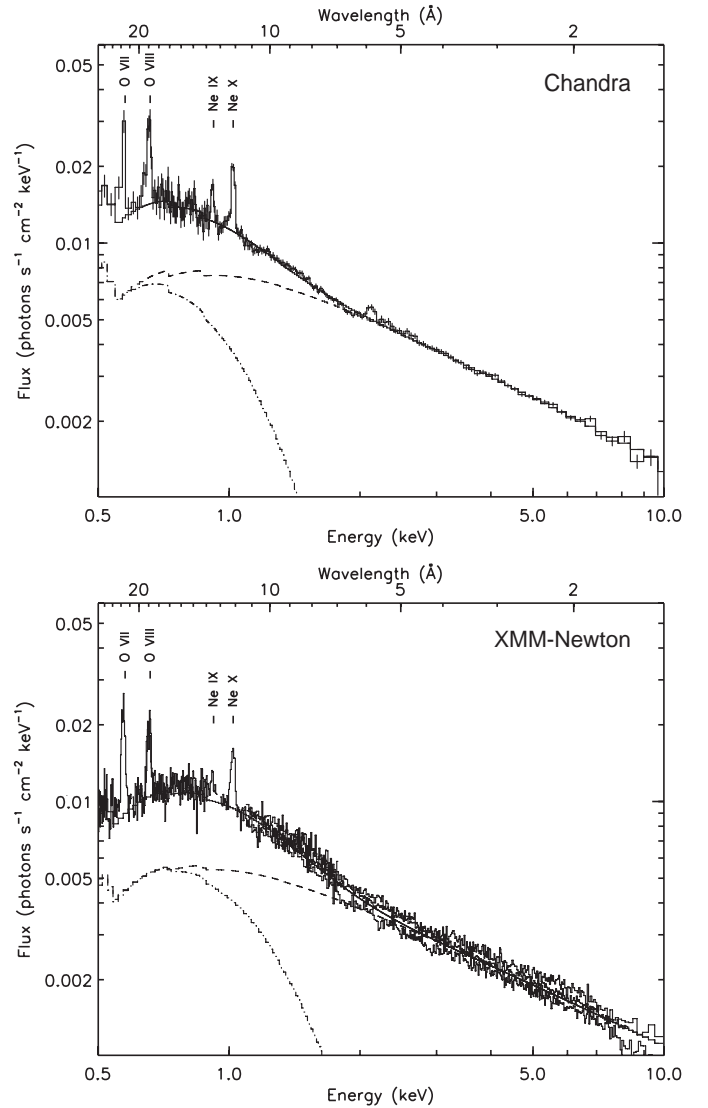}}
	\caption{The combined \chandra~(top panel) and \xmm~(bottom panel) spectra with representative continuum models comprised of absorbed blackbody plus power-law emission.  The blackbody (dot-dashed line) and power-law (dashed line) contributions are also plotted.}
	\label{both_cont_plot}
\end{figure}

We grouped the MOS data to have a minimum of 50 counts and the PN data to have a minimum of 500 counts per bin, and the RGS data to contain 6 channels per bin (resulting in $\simeq$ 0.04 \AA~wavelength bins, comparable to the instrumental resolution).  We grouped the HEG and MEG spectra to contain 4 channels and a minimum of 10 counts per bin (resulting in $\gtrsim$ 0.01 [0.02] \AA~wavelength bins for the HEG [MEG] data, again comparable to the instrumental resolution).

\begin{deluxetable*}{lcccccccc}
\tablewidth{0pt}
\tabletypesize{\scriptsize}
\tablecaption{Continuum Spectral Fits}
\tablehead{
\colhead{} & \colhead{} & \multicolumn{2}{c}{Power-law} & \colhead{} & \multicolumn{2}{c}{Blackbody} & \colhead{} \\
\cline{3-4} \cline{6-7}
\colhead{} & \colhead{$N_{\rm H}$} & \colhead{} & \colhead{} & \colhead{} & \colhead{} & \colhead{$kT$} & \colhead{} & \colhead{} \\
\colhead{Observation} & \colhead{($10^{21} \rm cm^{-2}$)} & \colhead{Norm\tablenotemark{1}} & \colhead{$\Gamma$} & \colhead{} & \colhead{$R^{2}_{\rm km}/D^{2}_{10~\rm kpc}$} & \colhead{(keV)} & \colhead{Flux\tablenotemark{2}} & \colhead{$C_{\nu}$ (dof)\tablenotemark{3}}
}
\startdata
\chandra~ObsID 104 &  $1.3^{+0.4}_{-0.3}$ & $12.1\pm0.5$ & $0.88\pm0.03$ & & $600^{+400}_{-200}$ & $0.21\pm0.02$ & 2.2 & 1.04 (1844) \\
{\it XMM} ObsID 0111070201 & $1.39^{+0.07}_{-0.09}$ & $8.0^{+0.1}_{-0.2}$ & $0.80\pm0.01$ & & $330^{+30}_{-40}$ & $0.254^{+0.008}_{-0.005}$ & 1.7 & 1.15 (1888) \\
\chandra~ObsID 3504 & $1.0^{+0.4}_{-0.3}$ & $8.4\pm0.2$ & $0.81\pm0.02$ & & $600^{+400}_{-200}$ & $0.19\pm0.01$ & 1.7 & 1.10 (1926) \\
 {\it XMM} ObsID 0152620101 & $1.38^{+0.06}_{-0.04}$ & $6.76^{+0.08}_{-0.07}$ & $0.782^{+0.008}_{-0.007}$ & & $290^{+30}_{-20}$ & $0.245^{+0.003}_{-0.005}$ & 1.5 & 1.38 (1857) \\
\enddata
\label{continuumFits}
\tablenotetext{1}{Normalization of power-law component at 1 keV in units of $10^{-3}$ \ergcmkev.}
\tablenotetext{2}{Absorbed 0.3--10.0 keV flux in units of $10^{-10}$ \ergcms.}
\tablenotetext{3}{The reduced Cash statistic and number of degrees of freedom for the fit.}
\end{deluxetable*}

Although a simple power-law model gives a reasonable fit ($\chi^{2}_{\nu} \simeq 1.2$), the residuals suggest the presence of an additional spectral component.  This has previously been modeled as a second power-law or a blackbody component (S01).  An F-test indicates that the addition of a blackbody component is preferred over a single power-law at the 8 $\sigma$ level, and this is what we use for our continuum fits.  To model the absorption, we used an updated version of the {\tt tbabs} model, {\tt tbnew\footnote{http://astro.uni-tuebingen.de/$\sim$wilms/research/tbabs/}} \citep{wilms07}, which includes high-resolution structure for the Ne K, Fe L, and O K edges.  The best-fit continuum models are presented in Table~\ref{continuumFits}, and the combined continuum spectra are shown in Figure~\ref{both_cont_plot}.

\begin{figure}
 \resizebox{3.5in}{!}{\includegraphics{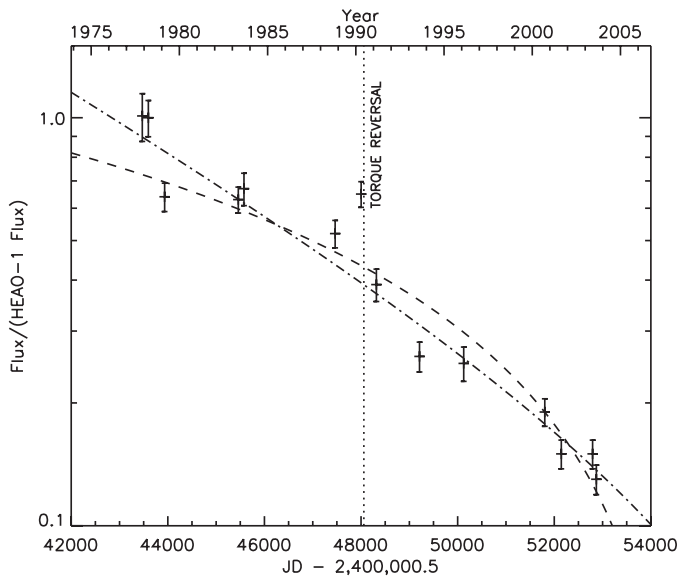}}
           \caption{The X-ray flux history of 4U 1626--67 from 1977-2003.  The dashed line is a linear fit; the dot-dash line a logarithmic fit to the data.  Errorbars represent the 1 $\sigma$ confidence intervals.  The most recent four data points represent the \chandra~and \xmm~observations.}
           \label{fluxhist}
\end{figure}

The 0.3--10 keV flux followed a decreasing trend, falling from $2.2 \times 10^{-10}~\ergcms$ in 2000 September to $1.5 \times 10^{-10}~\ergcms$ in 2003 August.  Overall, 4U 1626--67 has decreased in flux since 1977, with no apparent change following the reversal in accretion torque in 1990 (see Figure~\ref{fluxhist}). 

\subsection{Emission Lines}\label{emlines}

\begin{figure}
\resizebox{3.5in}{!}{\includegraphics{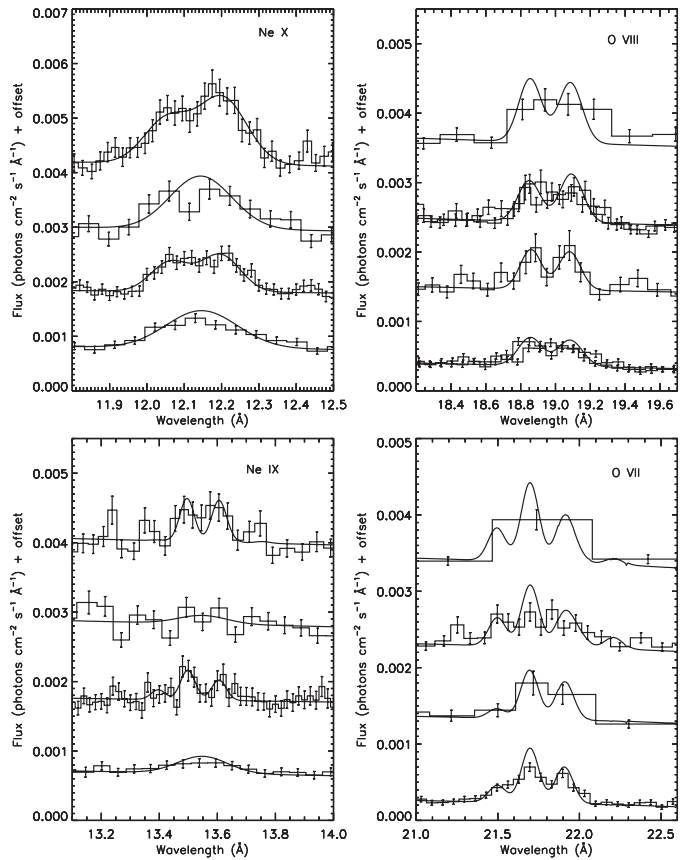}}
	\caption{Emission line regions for all four datasets.  The hydrogen-like \ion{Ne}{10} (left) and \ion{O}{8} (right) ions are in the top panels, and the helium-like triples of \ion{Ne}{9} (left) and \ion{O}{7} (right) are in the bottom panels. Each panel contains, from top to bottom, \chandra~ObsID 104, \xmm~ObsID 0111070201, \chandra~ObsID 3504, and \xmm~ObsID 0152620101.}
	\label{allLines}
\end{figure}

There are prominent Ne and O emission lines in all four spectra (see Figures~\ref{both_cont_plot} and~\ref{allLines}).  To explore the line characteristics, we added Gaussian features to the continuum models determined from the overall spectral fits.  All the lines are well-fit by single Gaussian profiles.  We note that the \ion{Ne}{9} lines are not reliably detected in the \xmm~data; for these lines, we fixed the line width and location to the \chandra~values to obtain approximations of the line flux.  The single-Guassian line fits are presented in Table~\ref{1gfits}.  

\begin{deluxetable*}{llcccccc}
\tablewidth{0pt}
\tabletypesize{\small}
\tablecaption{Single-Gaussian Emission Line Fits}
\tablehead{
\colhead{} & \colhead{} & \colhead{$\Delta\lambda$} & \colhead{$V$} & \colhead{FWHM} & \colhead{} & \colhead{EqW} & \colhead{} \\
\colhead{Instrument} & \colhead{MJD} & \colhead{(\AA)} & \colhead{(km s$^{-1}$)} & \colhead{(km s$^{-1}$)} & \colhead{Flux\tablenotemark{1}} & \colhead{(\AA~[eV])} & \colhead{$C_{\nu}$ (dof)\tablenotemark{2}}
}
\startdata
\multicolumn{8}{c}{Ne X line at 12.13 \AA} \\
\tableline
\chandra & 51803.6 & $0.021\pm0.014$  &  $ 510\pm340$ & $6330^{+ 860}_{- 690}$  &  $41.3^{+4.6}_{-4.3}$ & 0.29 [24]  &  0.97 (115)  \\ 
\xmm & 52145.1 & $0.016^{+0.033}_{-0.035}$  &  $ 390^{+ 830}_{- 880}$ & $4800^{+1800}_{-1600}$  &  $25.0^{+6.0}_{-5.7}$ & 0.21 [17]  &  1.22 (16)  \\ 
\chandra & 52795.1 & $0.018^{+0.012}_{-0.013}$  &  $ 440^{+ 310}_{- 320}$ & $5540^{+ 690}_{- 590}$  &  $19.4^{+2.2}_{-2.1}$ & 0.20 [17]  &  1.04 ( 134)  \\ 
\xmm & 52871.2 & $0.019^{+0.021}_{-0.022}$  &  $ 470^{+ 520}_{- 540}$ & $5540^{+1260}_{- 990}$  &  $20.4^{+3.0}_{-2.5}$ & 0.21 [18]  &  2.15 (16)  \\ 
\tableline
\multicolumn{8}{c}{Ne IX line at 13.55 \AA} \\
\tableline
\chandra & 51803.6 & $0.010^{+0.034}_{-0.032}$  &  $ 210^{+ 740}_{- 710}$ & $4200^{+1800}_{-1200}$  &  $13.2^{+4.5}_{-4.1}$ & 0.10 [6.8]  &  1.02 (91)  \\ 
\xmm\tablenotemark{3} & 52145.1 & \nodata  & \nodata & $4200$  &  $3.4^{+4.6}_{-3.4}$ & 0.030 [2.0]  &  1.03 (17)  \\ 
\chandra & 52795.1 & $-0.019^{+0.036}_{-0.039}$  &  $-430^{+ 800}_{- 850}$ & $4500^{+2000}_{-1700}$  &  $7.2^{+2.6}_{-2.5}$ & 0.079 [5.4]  &  0.90 ( 102)  \\ 
\xmm\tablenotemark{3} & 52871.2 & \nodata &  \nodata & $4500$  &  $7.1^{+2.4}_{-2.1}$ & 0.070 [4.7]  &  0.81 (17)  \\ 
\tableline
\multicolumn{8}{c}{O VIII line at 18.97 \AA} \\
\tableline
\chandra & 51803.6 & $0.063^{+0.058}_{-0.079}$  &  $1000^{+ 920}_{-1250}$ & $7100^{+3200}_{-1700}$  &  $60^{+17}_{-15}$ & 0.56 [19]  &  1.76 (45)  \\ 
\xmm & 52145.1 & $0.025^{+0.034}_{-0.032}$  &  $ 390^{+ 530}_{- 500}$ & $5510^{+1150}_{- 860}$  &  $39.1^{+6.6}_{-5.4}$ & 0.49 [17]  &  1.43 (60)  \\ 
\chandra & 52795.1 & $-0.016^{+0.059}_{-0.060}$  &  $-260\pm940$ & $6000^{+3400}_{-1600}$  &  $28.8^{+9.6}_{-8.2}$ & 0.40 [14]  &  0.99 (  56)  \\ 
\xmm & 52871.2 & $-0.006^{+0.022}_{-0.021}$  &  $ -100^{+ 350}_{- 340}$ & $5840^{+ 890}_{- 610}$  &  $26.1^{+2.8}_{-2.3}$ & 0.40 [14]  &  1.34 (  61)  \\ 
\tableline
\multicolumn{8}{c}{O VII line at 21.80 \AA} \\
\tableline
\chandra & 51803.6 & $-0.050^{+0.074}_{-0.077}$  &  $-688^{+1015}_{-1057}$ & $6154^{+1974}_{-1487}$  &  $77.3^{+26.3}_{-24.3}$ & 0.87 [23]  &  0.98 (22)  \\ 
\xmm & 52145.1 & $-0.027^{+0.070}_{-0.093}$  &  $-380^{+ 960}_{-1280}$ & $8900^{+4300}_{-1600}$  &  $72^{+18}_{-13}$ & 0.96 [25]  &  1.08 (50)  \\ 
\chandra & 52795.1 & $-0.020^{+0.093}_{-0.058}$  &  $-280^{+1280}_{- 800}$ & $3700^{+3600}_{-1400}$  &  $31^{+16}_{-12}$ & 0.50 [13]  &  0.91 (  25)  \\ 
\xmm & 52871.2 & $-0.051^{+0.022}_{-0.023}$  &  $-710^{+ 310}_{- 320}$ & $5200^{+ 850}_{- 490}$  &  $47.6^{+5.2}_{-4.4}$ & 0.86 [23]  &  1.18 (  51)  \\ 
\enddata
\label{1gfits}
\tablenotetext{1}{The Gaussian normalization in units of $10^{-5}~\rm photons~cm^{-2}~s^{-1}$. }
\tablenotetext{2}{The reduced Cash statistic and number of degrees of freedom for the fit.}
\tablenotetext{3}{Since the XMM data were not able to constrain the \ion{Ne}{9} lines, their positions and widths were fixed to the values found in the \chandra~data.}
\end{deluxetable*}

\begin{figure}
 \resizebox{3.3in}{!}{\includegraphics{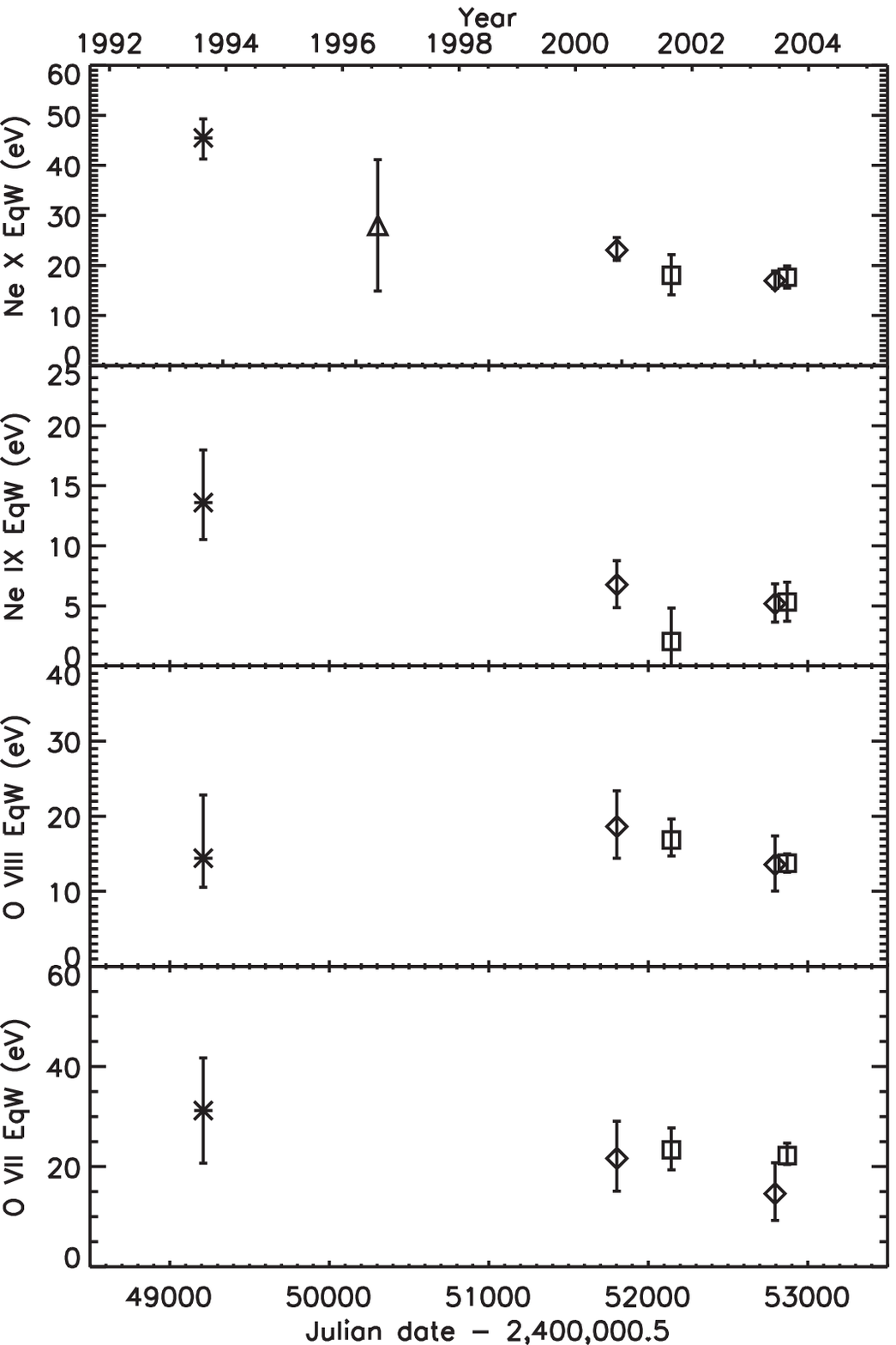}}
           \caption{Line equivalent widths from the literature and from the single-Gaussian fits presented in this paper.  The \asca~measurements are plotted with stars \citep{angelini95}, \bepposax~with a triangle \citep{owens97},  \chandra~with diamonds, and \xmm~with squares; error-bars represent the 90\% confidence intervals.}
           \label{eqws}
\end{figure}

The single-Gaussian fits reveal lines which are quite broad ($\approx$ 6000 km $\rm s^{-1}$ FWHM).  For a given line, the width does not vary significantly over the four observations.  However, the line strengths of \ion{Ne}{10} and \ion{O}{8} decrease over the course of the observations (the trend is not so clear for \ion{Ne}{9} or \ion{O}{7}).  To determine whether the line flux is decreasing more rapidly than the overall continuum flux, we fixed the line widths to their weighted averages and recalculated the line strengths, then fit the flux trends for the \ion{Ne}{10} and \ion{O}{8} lines.  Fixing the widths did not change the flux values appreciably, and improved the errors on the fluxes by only a small amount.  Over the $\approx 3$ year span of the observations, from 2000 September to 2003 August, the \ion{Ne}{10} flux decreased to $51.6\pm5.5\%$ and the \ion{O}{8} flux to $54.0\pm8.7\%$ of their initial values.  This decrease is significantly more than the decrease in the 0.3$-$10.0 keV continuum flux, which only fell to $76.4\pm6.6\%$ of its initial value.  The fact that the line strengths decrease faster than the continuum strength is also reflected in the declining equivalent widths of the lines.  We find that the equivalent widths of \ion{Ne}{10} and \ion{O}{8} decrease to $75.7\pm7.9\%$ and $74.3\pm12.3\%$ of their initial values, respectively (see Figure~\ref{eqws}, which also includes data from \asca~and \bepposax).

The {\it Chandra} \ion{Ne}{10} and \ion{O}{8} lines, as well as the \xmm~\ion{O}{7} line, show what appear to be double-peaked profiles (see Figure~\ref{allLines}).  As S01 point out, the line shapes, combined with the apparent high-density environment of the line-formation regions (as indicated by the dominance of the intercombination component of the He-like triplets; see below), suggest that the lines arise in an accretion disk.  To derive physically meaningful values for the line parameters, we fit each line with a pair of Gaussians, fixing the width and absolute velocity of the red- and blue-shifted components to be equal, thus approximating the expected disk-line ``two-horn'' profile.  However, there are a number of instances where it is not possible to constrain the width and velocity of a particular line.  For these instances, we fixed the values to the weighted averages from the observations where the line was well constrainted.  This allowed us to evaluate the approximate fluxes of the line components, with the assumption that the widths and velocities of a given line do not change substantially over the course of the observations.  Where multiple measurements could be made, this is true both for the widths from the single-Gaussian fits and for the widths and velocities from the double-Gaussian fits, so we consider it to be a reasonable assumption.  The \ion{Ne}{9} lines in the \xmm~data were too poorly constrained to be included in the double-Gaussian fits.  Table~\ref{2gfits} contains the values obtained from these fits.

\begin{deluxetable*}{llccccccccc}
\tablewidth{0pt}
\tabletypesize{\tiny}
\tablecaption{Double-Gaussian Emission Line Fits}
\tablehead{
\colhead{} & \colhead{} & \colhead{} & \colhead{} & \colhead{} & \multicolumn{2}{c}{Blueshifted Lines} & \colhead{} & \multicolumn{2}{c}{Redshifted Lines} & \colhead{} \\
\cline{6-7}  \cline{9-10}
\colhead{} & \colhead{} & \colhead{FWHM} & \colhead{$\Delta\lambda$} & \colhead{$V$} & \colhead{} & \colhead{EqW} & \colhead{} & \colhead{} & \colhead{EqW} & \colhead{} \\
\colhead{Observatory} & \colhead{MJD} & \colhead{(km s$^{-1}$)} & \colhead{(\AA)} & \colhead{(km s$^{-1}$)} & \colhead{Flux\tablenotemark{1}} & \colhead{(\AA~[eV])} & \colhead{} & \colhead{Flux\tablenotemark{1}} & \colhead{(\AA~[eV])} & \colhead{$C_{\nu}$ (dof)\tablenotemark{2}} 
}
\startdata
\multicolumn{11}{c}{Ne X line at 12.13 \AA} \\
\tableline
\chandra & 51803.6 & $3690^{+1190}_{- 630}$  &  $\pm0.075^{+0.011}_{-0.008}$ & $\pm1840^{+ 270}_{- 200}$ &  $15.8^{+3.1}_{-3.0}$ & 0.104 [8.84]  &  &  $23.8^{+3.8}_{-3.4}$ & 0.160 [13.36]  &  1.02 (246)  \\ 
\xmm & 52145.1 & $1500^{+1800}_{-1500}$  &  $\pm0.077^{+0.018}_{-0.017}$ & $\pm1910^{+ 450}_{- 430}$  &  $11.1^{+4.8}_{-4.6}$ & 0.092 [7.82]  & &  $14.4^{+5.2}_{-4.7}$ & 0.120 [9.99]  &  1.15 (33)  \\ 
\chandra & 52795.1 & $3040^{+ 670}_{- 440}$  &  $\pm0.068\pm0.007$ & $\pm1670^{+ 180}_{- 170}$   &  $8.2^{+1.5}_{-1.4}$ & 0.083 [7.05]  & &  $10.5^{+1.8}_{-1.6}$ & 0.107 [8.94]  & 1.10 (281)  \\ 
\xmm\tablenotemark{3} & 52871.2  & $3085$  &  $\pm0.072^{+0.017}_{-0.015}$ & $\pm1780^{+ 420}_{- 380}$  &  $11.3^{+2.6}_{-2.6}$ & 0.113 [9.45]  & &  $9.0^{+2.5}_{-2.6}$ & 0.089 [7.60]  &  1.86 (34)  \\ 
\tableline

\multicolumn{11}{c}{Ne IX resonance line at 13.45 \AA} \\
\tableline
\chandra\tablenotemark{4} & 51803.6 & $1300$  &  $\pm0.055$ & $\pm1210$  &  $0.1^{+1.7}_{-0.1}$ & 0.001 [0.05]  & &  $0.1^{+1.7}_{-0.1}$ & 0.001 [0.05]  &  1.09 (113)  \\ 
\chandra & 52795.1 & $1300^{+ 960}_{- 610}$  &  $\pm0.055^{+0.017}_{-0.018}$ & $\pm1210^{+ 380}_{- 400}$  &  $1.1^{+1.0}_{-1.0}$ & 0.012 [0.80]  & &  $1.1^{+1.0}_{-1.0}$ & 0.012 [0.80]  &  0.86 (120)  \\ 
\tableline

\multicolumn{11}{c}{Ne IX intercombination line at 13.55 \AA} \\
\tableline
\chandra\tablenotemark{4} & 51803.6 & $1300$  &  $\pm0.055$ & $\pm1210$  &  $5.0^{+1.8}_{-1.9}$ & 0.037 [2.52]  & &  $5.0^{+1.8}_{-1.9}$ & 0.037 [2.52]  &  1.09 (113)  \\ 
\chandra & 52795.1 & $1300^{+ 960}_{- 610}$  &  $\pm0.055^{+0.017}_{-0.018}$ & $\pm1210^{+ 380}_{- 400}$  &  $2.3^{+1.2}_{-1.1}$ & 0.021 [1.46]  & &  $2.3^{+1.2}_{-1.1}$ & 0.021 [1.46]  &  0.86 (120)  \\ 
\tableline

\multicolumn{11}{c}{Ne IX forbidden line at 13.70 \AA} \\
\tableline
\chandra\tablenotemark{4} & 51803.6 & $1300$  &  $\pm0.055$ & $\pm1210$  &  $0.2^{+1.7}_{-0.2}$ & 0.001 [0.09]  & &  $0.2^{+1.7}_{-0.2}$ & 0.001 [0.09]  &  1.09 (113)  \\ 
\chandra & 52795.1 & $1300^{+ 960}_{- 610}$  &  $\pm0.055^{+0.017}_{-0.018}$ & $\pm1210^{+ 380}_{- 400}$  &  $0.0^{+1.2}_{-0.0}$ & 0.0 [0.0]  & &  $0.0^{+1.2}_{-0.0}$ & 0.0 [0.0]  &  0.86 (120)  \\ 
\tableline

\multicolumn{11}{c}{O VIII line at 18.97 \AA} \\
\tableline
\chandra\tablenotemark{5} & 51803.6 & $2474$  &  $\pm0.116$ & $\pm1840$  &  $26.3^{+8.7}_{-9.9}$ & 0.258 [8.78]  & &  $21.6^{+10.8}_{-6.4}$ & 0.210 [7.31]  &  1.72 (50)  \\
\xmm & 52145.1 & $2450^{+ 900}_{- 610}$  &  $\pm0.122^{+0.016}_{-0.017}$ & $\pm1930\pm260$  &  $17.6^{+4.7}_{-4.4}$ & 0.217 [7.58]  & &  $21.9^{+4.9}_{-4.5}$ & 0.275 [9.36]  &  1.10 (73)  \\ 
\chandra & 52795.1 & $2320^{+1140}_{- 650}$  &  $\pm0.112^{+0.021}_{-0.023}$ & $\pm1770^{+ 330}_{- 370}$  &  $13.0^{+5.7}_{-4.9}$ & 0.179 [6.24]  & &  $13.7^{+5.8}_{-5.0}$ & 0.191 [6.51]  &  0.98 (61)  \\ 
\xmm & 52871.2 & $2600^{+ 870}_{- 600}$  &  $\pm0.115^{+0.012}_{-0.012}$ & $\pm1810^{+ 180}_{- 200}$  &  $12.7^{+2.1}_{-1.9}$ & 0.191 [6.66]  & &  $12.4^{+2.0}_{-1.8}$ & 0.189 [6.43]  &  1.15 (76)  \\ 
\tableline

\multicolumn{11}{c}{O VII resonance line at 21.60 \AA} \\
\tableline
 \chandra\tablenotemark{6} & 51803.6 & $1750$  &  $\pm0.104$ & $\pm1430$  &  $12.8^{+11.2}_{-8.9}$ & 0.142 [3.82]  & &  $12.8^{+11.2}_{-8.9}$ & 0.142 [3.82]  &  1.16 (22)  \\
\xmm\tablenotemark{6} & 52145.1 & $1750$  &  $\pm0.104$ & $\pm1430$  &  $11.7^{+5.4}_{-5.2}$ & 0.171 [4.60]  & &  $11.7^{+5.4}_{-5.2}$ & 0.171 [4.60]  &  1.16 (47)  \\
\chandra\tablenotemark{6} & 52795.1 & $1750$  &  $\pm0.104$ & $\pm1430$  &  $3.1^{+7.0}_{-3.1}$ & 0.048 [1.30]  & &  $3.1^{+7.0}_{-3.1}$ & 0.048 [1.30]  &  0.95 (25)  \\ 
\xmm & 52871.2 & $1750 ^{+ 560}_{- 440}$  &  $\pm0.104^{+0.016}_{-0.016}$ & $\pm1430\pm220$  &  $6.9^{+2.3}_{-0.6}$ & 0.122 [3.27]  & &  $6.9^{+2.3}_{-0.6}$ & 0.122 [3.27]  &  0.96 (49)  \\ 
\tableline

\multicolumn{11}{c}{O VII intercombination line at 21.80 \AA} \\
\tableline
 \chandra\tablenotemark{6} & 51803.6 & $1750$  &  $\pm0.104$ & $\pm1430$  &  $19.8^{+12.6}_{-11.0}$ & 0.110 [2.89]  & &  $19.8^{+12.6}_{-11.0}$ & 0.110 [2.89]  &  1.16 (22)  \\
\xmm\tablenotemark{6} & 52145.1 & $1750$  &  $\pm0.104$ & $\pm1430$  &  $15.3^{+5.5}_{-7.1}$ & 0.100 [2.63]  & &  $15.3^{+5.5}_{-7.1}$ & 0.100 [2.63]  &  1.16 (47)  \\
\chandra\tablenotemark{6} & 52795.1 & $1750$  &  $\pm0.104$ & $\pm1430$  &  $13.6^{+6.9}_{-7.2}$ & 0.159 [4.19]  & &  $13.6^{+6.9}_{-7.2}$ & 0.159 [4.19]  &  0.95 (25)  \\ 
\xmm & 52871.2 & $1750^{+ 560}_{- 440}$  &  $\pm0.104^{+0.016}_{-0.016}$ & $\pm1430 \pm220$  &  $16.6^{+1.4}_{-2.6}$ & 0.156 [4.11]  & &  $16.6^{+1.4}_{-2.6}$ & 0.156 [4.11]  &  0.96 (49)  \\ 
\tableline

\multicolumn{11}{c}{O VII forbidden line at 22.10 \AA} \\
\tableline
 \chandra\tablenotemark{6} & 51803.6 & $1750$  &  $\pm0.104$ & $\pm1430$  &  $2.8^{+10.0}_{-2.8}$ & 0.020 [0.52]  & &  $2.8^{+10.0}_{-2.8}$ & 0.020 [0.52]  &  1.16 (22)  \\
\xmm\tablenotemark{6} & 52145.1 & $1750$  &  $\pm0.104$ & $\pm1430$  &  $5.7^{+5.3}_{-3.8}$ & 0.057 [1.46]  & &  $5.7^{+5.3}_{-3.8}$ & 0.057 [1.46]  &  1.16 (47)  \\
\chandra\tablenotemark{6} & 52795.1  & $1750$  &  $\pm0.104$ & $\pm1430$  &  $0.0^{+2.7}_{-0.0}$ & 0.0 [0.0]  & &  $0.0^{+2.7}_{-0.0}$ & 0.0 [0.0]  &  0.95 (25)  \\ 
\xmm & 52871.2 & $1750 ^{+ 560}_{- 440}$  &  $\pm0.104^{+0.017}_{-0.016}$ & $\pm1430 \pm220$  &  $0.0^{+2.2}_{-0.0}$ & 0.0 [0.0]  & &  $0.0^{+2.2}_{-0.0}$ & 0.0 [0.0]  &  0.96 (49)  \\ 

\enddata
\tablenotetext{1}{The Gaussian normalization in units of $10^{-5}~\rm photons~cm^{-2}~s^{-1}$. }
\tablenotetext{2}{The reduced Cash statistic and number of degrees of freedom for the fit.}
\tablenotetext{3}{The Ne X line width was not well constrained for this observation, and its value was fixed to the weighted average of the values found in the other observations.}
\tablenotetext{4}{The Ne IX line widths and velocities were not well constrained for this observation and were fixed to the values found in the \chandra~observation of MJD 52795.1.}
\tablenotetext{5}{The O VIII line width and velocity were not well constrained for this observation and were fixed to the weighted averages of the values found in the other observations.}
\tablenotetext{6}{The O VII line width and velocity were not well constrained for this observation and were fixed to the values found in the \xmm~observation of MJD 52871.2.}
\label{2gfits}
\end{deluxetable*}

Fitting the helium-like triplets of \ion{Ne}{9} and \ion{O}{7} with multiple Gaussians allowed us to estimate the strengths of the resonance, intercombination, and forbidden lines.  The \ion{Ne}{9} intercombination line is clearly dominant in both \chandra~observations (we were unable to fit this line in the \xmm~observations).  While the intercombination line is the strongest of the \ion{O}{7} triplet as well, the resonance line contains significant flux, whereas the forbidden lines do not contribute substantially to any of the triplets.  The lack of forbidden lines suggests that we are observing emission from a high-density plasma ($\rm n_{e} \gtrsim 10^{12}~cm^{-3}$), which in turn supports the hypothesis that the lines arise in the accretion disk.  

\subsection{Photoelectric Absorption and Elemental Abundances}\label{photabs}

\begin{deluxetable}{llccc}
\tablewidth{0pt}
\tabletypesize{\tiny}
\tablecaption{4U 1626--67 absorption edge fits}
\tablehead{
\colhead{$\lambda$} & \colhead{Edge} & \colhead{Observatory} & \colhead{Total $N_{\rm H}$\tablenotemark{1}} & \colhead{Implied $N_{\rm Z}$\tablenotemark{2}} \\
\colhead{(\AA)} & \colhead{} & \colhead{(ObsID)} & \colhead{($\rm 10^{21}~cm^{-2}$)} & \colhead{($\rm 10^{17}~cm^{-2}$)} }
\startdata
14.3 & Ne K & \chandra~(104) & $3.0^{+1.1}_{-2.5}$ & $2.6^{+0.9}_{-2.2}$ \\
	&	   & {\it XMM}~(0111070201) & $1.9^{+0.2}_{-1.1}$ & $1.67^{+0.17}_{-0.95}$ \\	
         &          & \chandra~(3504) &  $0.9^{+2.2}_{-0.5}$ & $0.8^{+1.9}_{-0.4}$ \\
	&	   & {\it XMM}~(0152620101) &  $1.20^{+0.45}_{-0.33}$ & $1.05^{+0.40}_{-0.29}$ \\
\tableline
17.5 & Fe L$_{\rm III}$ & \chandra~ObsID 104 & $0.6^{+2.1}_{-0.3}$ & $0.17^{+0.56}_{-0.07}$ \\
	&	   & {\it XMM}~(0111070201) & $1.38^{+0.70}_{-0.17}$ & $0.37^{+0.19}_{-0.05}$ \\	
         &                             & \chandra~(3504) & $0.5^{+1.8}_{-0.2}$ & $0.13^{+0.48}_{-0.05}$ \\
	&	   & {\it XMM}~(0152620101) & $1.29^{+0.35}_{-0.18}$ & $0.348^{+0.094}_{-0.049}$ \\
\tableline
23.3 & O K & \chandra~ObsID 104 & $1.2^{+1.5}_{-0.7}$ & $5.9^{+7.2}_{-3.4}$ \\
	&	   & {\it XMM}~(0111070201) & $1.38^{+0.42}_{-0.12}$ & $6.7^{+2.1}_{-0.6}$ \\	
         &         & \chandra~(3504) & $1.7^{+0.6}_{-1.4}$ & $8.4^{+3.0}_{-6.9}$ \\
	&	   & {\it XMM}~(0152620101) & $1.38^{+0.25}_{-0.04}$ & $6.8^{+1.2}_{-0.2}$ \\
\enddata
\tablenotetext{1}{Fitted value of $N_{\rm H}$, using the {\tt tbnew} model \citep{wilms07} and fitting over the immediate range of the absorption structure, restricting the continuum components to remain within their 90\% confidence intervals (see text).}
\tablenotetext{2}{Value of $N_{\rm Z}$ implied by the fitted $N_{\rm H}$, assuming ISM abundances presented in \citet{wilms00}.}
\label{edge_fits}
\end{deluxetable}

Since 4U 1626--67 shows prominent Ne and O emission lines, it is natural to ask whether this is a reflection of an overabundance of heavy elements in the system, or of the particular geometry and conditions inherent to this unique source.  One way to constrain the local abundances is by measuring the strength of the absorption complexes within our spectral range.  For the \xmm~and \chandra~data this includes the Ne K , Fe L, and O K edges.  We fit the edges over the wavelength ranges 13.8--14.8, 16.5--18.0, and 22.2--25.3 \AA, respectively, and used the {\tt tbnew} model, which includes high-resolution modeling of the edge cross-sections \citep{wilms07,juett04,juett06}.  In order to constrain the continuum over the restricted wavelength ranges, we used the same continuum model as was fitted to the full spectrum, restricting the model parameters to vary within the previously determined 90\% confidence intervals.  We present the fit results in Table~\ref{edge_fits}.  Although our results suggest a slight overabundance of these elements relative to the expected interstellar value of $N_{\rm H} = (6.2\pm0.7) \times 10^{20} \rm~cm^{-2}$ \citep[measured directly from the Ly$\alpha$ line;][]{wang02}, neither the \chandra~nor the \xmm~datasets require a statistically significant local contribution to the interstellar column depth (see Section~\ref{photabs_disc} for further discussion). 

\section{Timing Analysis}\label{timing_anal}

\begin{figure}
\resizebox{3.4in}{!}{\includegraphics{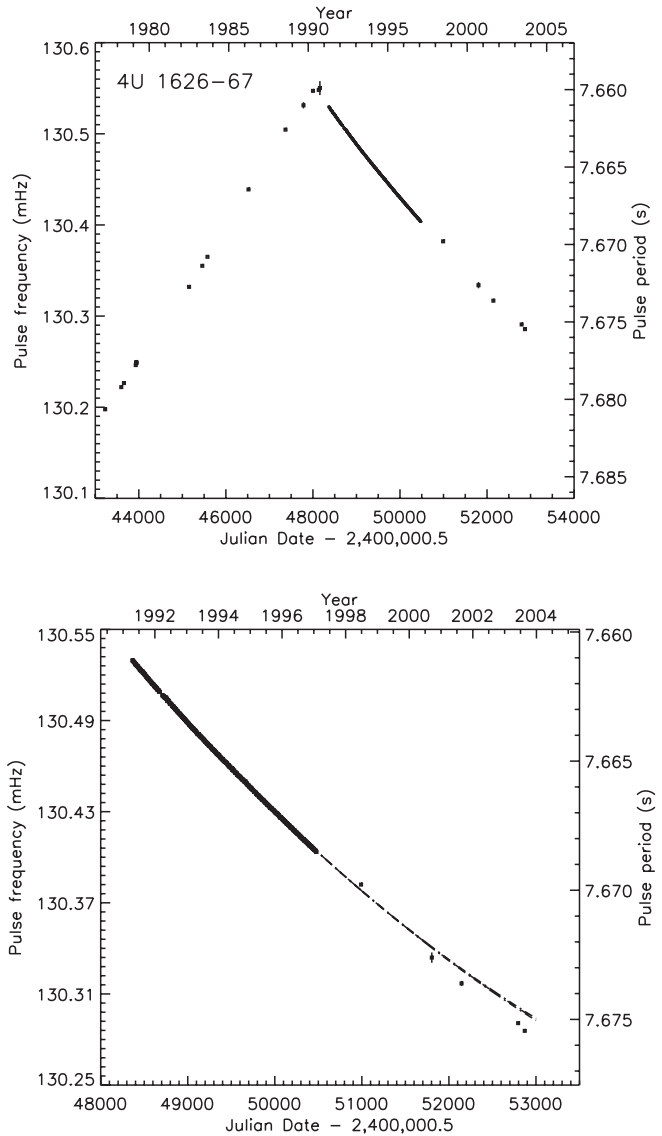}}
	\caption{Frequency history of 4U 1626$-$67.  Top panel shows all data; bottom panel is an expanded view of the {\it Compton}/BATSE through the \chandra~and \xmm~observations, along with the BATSE prediction (solid curve with dot-dash errors).  Pulse frequency data from other observatories, as well as the BATSE prediction, can be found in \citet{chakrabarty97} and \citet{chakrabarty01}.}
	\label{freqHist}
\end{figure}

All of the \chandra~and \xmm~observations of 4U 1626--67 have high enough time resolution to determine the pulse period, and the \xmm~observations also allowed us to create energy-resolved pulse profiles.  Prior to performing the timing analysis, we corrected the photon arrival times to the location of the solar system barycenter.  

\begin{figure}
\resizebox{3.3in}{!}{\includegraphics{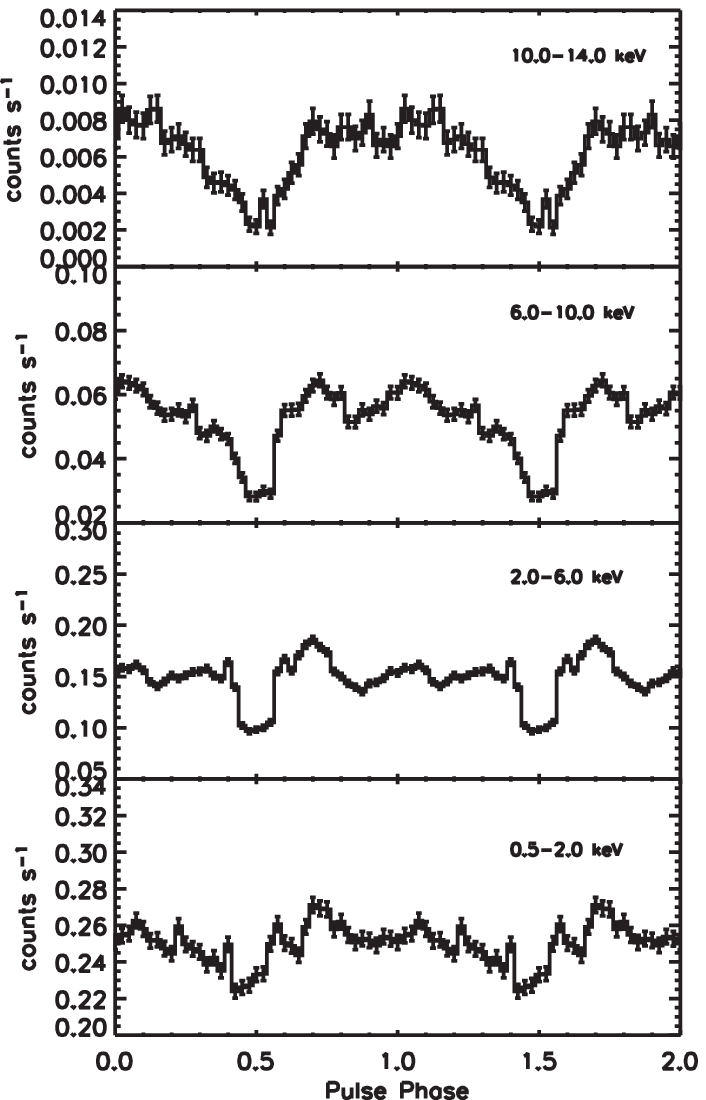}}
	\caption{Folded pulse profiles of 4U 1626$-$67 from the \xmm~PN, ObsID 0111070201.}
	\label{pulseProf011}
\end{figure}

\begin{figure}
\resizebox{3.3in}{!}{\includegraphics{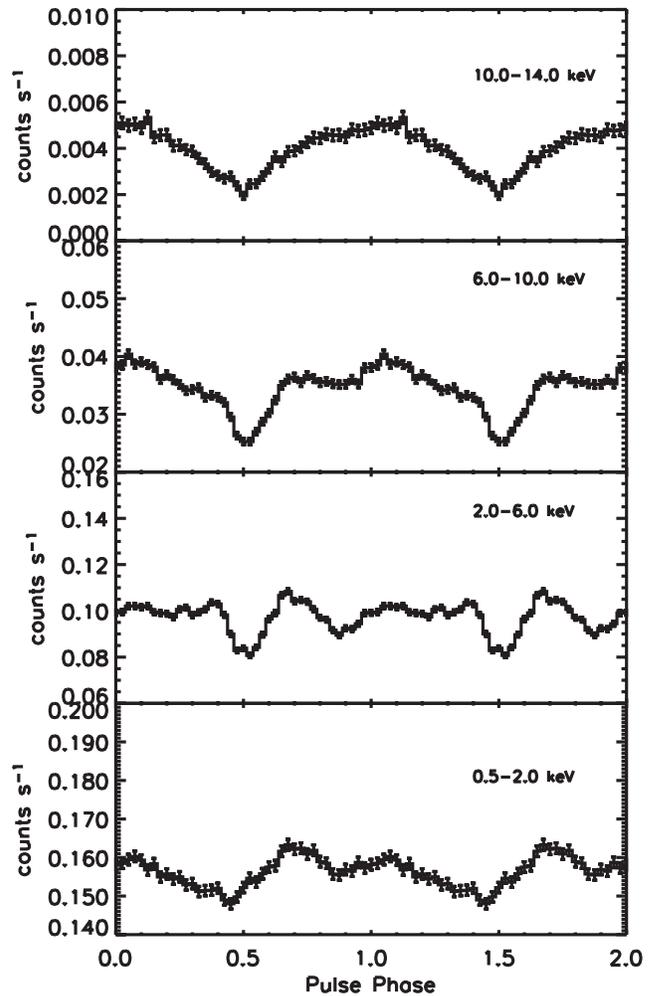}}
	\caption{Folded pulse profiles of 4U 1626$-$67 from the \xmm~PN, ObsID 0152620101.}
	\label{pulseProf015}
\end{figure}

To determine the pulse period, we used an oversampled fast Fourier transform, with the frequency and frequency error determined by the following formulae \citep{chakrabarty96, middleditch76a, middleditch76b}:

\begin{equation}
\hat{\nu}_{0} = \nu_{k} + \frac{3}{4\pi^{2}\epsilon T} \left(\frac{\bar{P}_{k+\epsilon} - \bar{P}_{k-\epsilon}}{\bar{P}_{k}}\right);
\end{equation}
\begin{equation}
\sigma_{\hat{\nu}_{0}} \approx \frac{1}{2\pi T} \sqrt{\frac{6}{\bar{P}_{k}}};
\end{equation}

where $\bar{P}$ is the normalized power, the $k^{th}$ bin contains the peak power, $T$ is the length of the observation, and $n = \epsilon^{-1}$ is the oversampling factor (for our analysis we used $n=6$). The results are presented in Table~\ref{timing}, and Figure~\ref{freqHist} shows a plot of the frequencies as well as the prediction curve and associated errors extrapolated from previous {\it Compton}/BATSE monitoring results \citep{chakrabarty97}.  As can be seen from the plot, the pulsar is spinning down faster than was predicted, which is significant with respect to the stability of the timing during the BATSE era, but not unexpected considering the variable spin history of the source.  We note that the QPO at 0.048 Hz is present in each of the observations, as has been seen previously \citep{shinoda90,chakrabarty98a}.

The XMM PN data are ideal for creating energy-resolved pulse profiles.  We made four energy cuts and divided the pulse period into 40 bins to create the profiles presented in Figures~\ref{pulseProf011} and \ref{pulseProf015}.  The profiles have some significant differences from what was seen previously \citep[see, e.g.][]{pravdo79,kii86,mcclintock80,levine88}.  In the earlier observations, the pulse profile shows a dip at energies $\lesssim 2$ keV, then the emergence of two prominent peaks which flank this dip as the energy increases, reaching a maximum at $\approx 5$ keV.  By $\approx 13$ keV, the dip has returned, and becomes both broader and deeper as the energy increases.  We find that the current pulse profiles lack the double-peak feature that was seen before, and show only a dip which broadens at lower and higher energies.  The profile is not entirely symmetric; there also appears to be a secondary dip at phase $\approx 0.85$.  Insofar as the pulse profile reflects the geometric and radiative properties of the accretion column and neutron star hot spot \citep[see, e.g.][]{kii86}, changes in the pulse profile suggest that there have been fundamental changes in these regions of the system.  While we do not speculate here as to the exact nature of these changes, we note that the accretion geometry may have changed at the time of torque reversal, affecting the accretion flow at the neutron star surface.  Therefore, the changes that are seen in the pulse profile may be related to the change in the continuum spectrum of 4U 1626$-$67 that was observed around the time of the torque reversal \citep{angelini95}.

We also note that there were no flaring events seen in any of our observations, in contrast to what has been previously observed, when 4U 1626$-$67 was seen to flare dramatically in both X-ray and optical data on timescales of $\approx 1000$ s \citep{joss78, mcclintock80, li80}.  The cessation of flaring activity may have occurred at the same time as the torque reversal, though it is not clear what physical mechanism was responsible for the flaring or why it would be correlated with the change in sign of $\dot{P}$ \citep{chakrabarty01}.

\begin{deluxetable}{llll}
\tablewidth{0pt}
\tabletypesize{\small}
\tablecaption{Pulse period of 4U 1626$-$67}
\tablehead{
 & & \colhead{Pulse Period} & \\
\colhead{Date} & \colhead{MJD} & \colhead{(s)} & \colhead{Observatory}}
\startdata
2000 Sep 16 & 51803.6 & 7.6726(2) & \chandra \\
2001 Aug 24 & 52145.1 & 7.6736(2) & \xmm \\
2003 Jun 5 & 52795.1 & 7.67514(5) & \chandra \\
2003 Aug 20 & 52871.2 & 7.67544(6) & \xmm \\
\enddata
\label{timing}
\end{deluxetable}

\section{Discussion}\label{disc}

\subsection{Absorption Edge Measurements}\label{photabs_disc}

Four known or candidate ultracompact binary systems have X-ray spectra which show evidence of high Ne/O ratios \citep[4U 0614+091, 2S 0918--549, 4U 1543--624, and 4U 1850--087;][]{juett01,juett03b}, though none of these systems have the X-ray emission line complexes seen in 4U 1626--67.  These unusually strong lines, which arise from highly ionized species of Ne and O, suggest that the donor must contain a significant quantity of these elements, and we have previously suggested that the donor may be the chemically fractionated core of a C-O-Ne or O-Ne-Mg white dwarf (S01).  However, since the strength of the emission lines is very dependent on local plasma parameters, it is difficult to use them to determine elemental abundances.  Alternatively, a way of determining local abundances is to measure the depths of the neutral absorption edges and to compare these with the value of the interstellar absorption along the line of sight (for example, as determined by the broader continuum fit).  Any absorption in excess of what is expected from standard ISM abundances could be taken as an indication of material local to the system.

The Ne K, Fe L, and O K absorption edges do not appear strong in any of our observations, including the first \chandra~observation (ObsID 104) that we analyzed in S01.   Therefore, we are no longer able to argue for a significant overabundance of O or Ne in cool circumstellar material as we suggested in S01, at least not on the grounds of the absorption edge analysis (see also Section~\ref{photabs}). Perhaps these elements are, in fact, overabundant in the disk (and thus the donor), but not much material is expelled from the system, or subsequently cools to form an absorbing medium.  Our results agree with a recent study of the optical spectra of 4U 0614+091 and 4U 1626--67 \citep{werner06}.  These spectra did not reveal any Ne emission features, only lines from C and O, which allowed them to place a limit on the Ne/O ratio of $\approx 0.2$.  Furthermore, 4U 0614+091 showed evidence of a higher C/O ratio than 4U 1626--67, suggesting that its donor is actually more evolved.  If this is the case, than the most significant difference between the two systems---the high magnetic field of 4U 1626--67--- could be the reason that 4U 1626--67 has prominent emission line features whereas the other ultracompact sources do not.

\subsection{Emission Line Characteristics and Physical Implications}

The emission lines have double-peaked profiles, and the line ratios imply that they are formed in a high-density environment, which suggests that they arise somewhere in the accretion disk (see also S01).  If they are, in fact, disk features, then the measured velocities correspond to the actual disk velocities by the relation 

\begin{equation}
\label{linev}
v_{\rm obs} = v \sin{i} = \sqrt{\frac{GM_{\rm x}} {r}} \sin{i}.
\end{equation}

The maximum disk velocity will occur at the inner edge of the accretion disk, which is truncated at the co-rotation radius of the neutron star at $\approx 6.5 \times 10^{8} \rm~cm$.  The Keplerian velocity at this radius is $\simeq 5400 \rm~km~s^{-1}$, whereas the line velocities are measured to be $\simeq 1700 \rm~km~s^{-1}$ (note that this is the velocity at the center of the Gaussian in the double-Gaussian fits, which is therefore an underestimate of the absolute maximum velocity).  If we take this to be the velocity at the inner disk edge, we are able to place a constraint on the inclination angle of the system: $\sin{i} \gtrsim 1700/5400 \simeq 0.31$ ($i \gtrsim 22 \degr$).  We may combine this with the previously derived upper limit on the projected semi-major axis, given an orbital period of 42 min and limits on the timing noise \citep{chakrabarty97}, to find $3 \lesssim a_{\rm x} \sin{i} \lesssim 8$ lt-ms.  

The line ratio measurements are indicators of the plasma conditions in the line formation regions.  Although we are not able to derive robust limits for the standard diagnostics, we note that the lack of detectable forbidden lines implies that the region is relatively high-density ($n_{\rm e} \gtrsim 10^{12} \rm~cm^{-3}$).  Furthermore, the relatively low value of the intercombination line ($i$) with respect to the resonance line ($r$), $r/i \approx 2$, suggests that the temperature of the line-formation region is $\gtrsim 10^{6}$ K \citep{porquet00}.  As was discussed in S01, this temperature is characteristic of the highly ionized, optically thin outer layers of the accretion disk.

\subsection{Long-term Flux Evolution}

X-ray observations of 4U 1626--67 show that it has been decreasing in flux since 1977.  Since this decrease appears to be bolometric---there is no sign of variable absorption, and the continuum spectrum is known to have undergone only one major change---we may use the X-ray flux to help determine other characteristics of the system.  The long-term trend can be described by either an exponential decay or a linear decrease (see Figure~\ref{fluxhist}).  If we take it to be logarithmic, and integrate over the duration of the outburst, the fluence is 0.927 erg cm$^{2}$ (giving a total energy of $1.1 \times 10^{44}~(d/1 \rm~kpc)^{2} \rm~erg$).  Converting this to an average accretion rate, we find that 

\begin{equation}
\label{aveacc}
\dot{M}_{\rm ave} \approx 2.2 \times 10^{-10}~\Delta t_{\rm yr}^{-1}~d_{\rm kpc}^{2}~M_{\sun}~\rm yr^{-1},
\end{equation}

where $\Delta t_{\rm yr}$ is the time between outbursts in years and $d_{\rm kpc}$ is the distance to the source in kpc.  If we assume that the source is persistent and take $\Delta t \approx 30$ yr, we find that $d \lesssim 2$ kpc, which is less than the minimum distance of 3~kpc derived from considerations of the accretion torque during spin-up \citep{chakrabarty97}.  If we assume a distance of $\gtrsim 3$ kpc, and take the long-term average accretion rate to be $3 \times 10^{-11}~M_{\sun} \rm~yr^{-1}$ \citep[the value from gravitational-radiation--driven orbital evolution with a white dwarf donor; see][]{chakrabarty98a}, we find that $\Delta t \gtrsim 70$ yr, significantly longer than the known duration of the current outburst.  

We may also combine current theory on disk stability \citep[see, e.g.][]{king96,vanparadijs96} with the flux data for 4U 1626--67 to obtain constraints on its distance and possible outburst duration.  If the effective disk temperature is that of an X-ray heated accretion disk, then

\begin{equation}
T^{4} = \frac{\eta \dot{M} c^2 (1 - \beta)} {4 \pi \sigma r^{2}} \left(\frac{dH} {dr} - \frac{H} {r}\right).
\end{equation}

Following \citet{king96}, we take $\eta = 0.11, \beta = 0.9$, assume that $H/r \approx 0.2$ is constant, and that $H \propto r^{n}$, where $n = 9/8$ for shallow heating, which we would expect to dominate in this system.  The disk stability criterion is that the temperature at the outermost disk radius remain $\gtrsim 6500$ K \citep{smak83}.  Setting $r = r_{\rm out} \simeq 2 \times 10^{10} \rm~cm$, we derive a critical accretion rate of $\dot{M} = 3.6 \times 10^{-12}~M_{\sun}~\rm yr^{-1}$.  Note that this is lower than $\dot{M}_{\rm GR}$.  Assuming that the accretion rate is directly proportional to the flux, and that it remained above the critical rate throughout the current outburst, we find the distance to 4U 1626--67 to be $\gtrsim 0.6$ kpc.  Furthermore, if we assume a distance range of 3--13 kpc and an exponential (or linear) decline in flux commensurate with the current trend, we find that $\dot{M} \lesssim \dot{M}_{\rm crit}$ in the range 2018--2020 (or 2008 June--Sept).  We therefore expect 4U 1626--67 to become quiescent within 2--15 years.  

\acknowledgments{We thank Mike Nowak, John Houck, Jake Hartman, and Dave Huenemoerder for their indispensable help and encouragement.  We are also grateful for the assistance of the folks at the {\it XMM} helpdesk, and of Mariano Mendez, with the reduction and analysis of the {\it XMM} data.  This research was supported by NASA through the contract to MIT for the \chandra/HETG instrument team as well as through the XMM grant NNG-04GL41G.}

\nocite{verbunt90}

\end{document}